\documentclass[twoside]{article}
\usepackage{fleqn,espcrc2}

\usepackage{graphicx}
\usepackage{epsfig}
\usepackage[figuresright]{rotating}

\begin{document}

%
%
\newcommand{\s}{\mbox{$s$}}
\newcommand{\ttra}{\mbox{$t$}}
\newcommand{\modt}{\mbox{$|t|$}}
\newcommand{\eminpz}{\mbox{$E-p_z$}}
\newcommand{\eminpzs}{\mbox{$\Sigma(E-p_z)$}}
\newcommand{\rap}{\ensuremath{\eta^*} }
\newcommand{\W}{\mbox{$W$}}
\newcommand{\w}{\mbox{$W$}}
\newcommand{\Q}{\mbox{$Q$}}
\newcommand{\q}{\mbox{$Q$}}
\newcommand{\xB}{\mbox{$x$}}  
\newcommand{\xF}{\mbox{$x_F$}}  
\newcommand{\xg}{\mbox{$x_g$}}  
\newcommand{\xbj}{$x$}
\newcommand{\xpom}{x_{I\!\!P}}
\newcommand{\zpom}{z_{I\!\!P}}
\newcommand{\y}{\mbox{$y~$}}
\newcommand{\Qsq}{\mbox{$Q^2$}}
\newcommand{\qzsq}{\mbox{$Q_o^2$}}
\newcommand{\qsq}{\mbox{$Q^2$}}
\newcommand{\kjet}{\mbox{$k_{T\rm{jet}}$}}
\newcommand{\xjet}{\mbox{$x_{\rm{jet}}$}}
\newcommand{\Ejet}{\mbox{$E_{\rm{jet}}$}}
\newcommand{\thjet}{\mbox{$\theta_{\rm{jet}}$}}
\newcommand{\pjet}{\mbox{$p_{T\rm{jet}}$}}
\newcommand{\et}{\mbox{$E_T$}}
\newcommand{\kt}{\mbox{$k_T$}}
\newcommand{\ptrans}{\mbox{$p_T~$}}
\newcommand{\pth}{\mbox{$p_T^h~$}}
\newcommand{\pte}{\mbox{$p_T^e~$}}
\newcommand{\ptsq}{\mbox{$p_T^{\star 2}~$}}
\newcommand{\as}{\mbox{$\alpha_s$}}
\newcommand{\ycut}{\mbox{$y_{\rm cut}~$}}
\newcommand{\gx}{\mbox{$g(x_g,Q^2)$~}}
\newcommand{\xpart}{\mbox{$x_{\rm part~}$}}
\newcommand{\mrsdm}{\mbox{${\rm MRSD}^-~$}}
\newcommand{\mrsdmp}{\mbox{${\rm MRSD}^{-'}~$}}
\newcommand{\mrsdn}{\mbox{${\rm MRSD}^0~$}}
\newcommand{\lambdams}{\mbox{$\Lambda_{\rm \bar{MS}}~$}}
%
%
\newcommand{\gp}{\ensuremath{\gamma}p }
\newcommand{\gammasp}{\ensuremath{\gamma}*p }
\newcommand{\gammap}{\ensuremath{\gamma}p }
\newcommand{\gsp}{\ensuremath{\gamma^*}p }
\newcommand{\dsiget}{\ensuremath{{\rm d}\sigma_{ep}/{\rm d}E_t^*} }
\newcommand{\dsigrap}{\ensuremath{{\rm d}\sigma_{ep}/{\rm d}\eta^*} }
\newcommand{\epem}{\mbox{$e^+e^-$}}
\newcommand{\ep}{\mbox{$ep~$}}
\newcommand{\epl}{\mbox{$e^{+}$}}
\newcommand{\emi}{\mbox{$e^{-}$}}
\newcommand{\epm}{\mbox{$e^{\pm}$}}
\newcommand{\xsec}{cross section}
\newcommand{\xsecs}{cross sections}
\newcommand{\inter}{interaction}
\newcommand{\inters}{interactions}
\newcommand{\eplp}{\mbox{$e^+p$}}
\newcommand{\emip}{\mbox{$e^-p$}}
\newcommand{\gamm}{$\gamma$}
\newcommand{\zzo}{$Z^{o}$}
%
%
\newcommand{\phib}{\mbox{$\varphi$}}
\newcommand{\rh}{\mbox{$\rho$}}
\newcommand{\rhz}{\mbox{$\rh^0$}}
\newcommand{\ph}{\mbox{$\phi$}}
\newcommand{\om}{\mbox{$\omega$}}
\newcommand{\jpsi}{\mbox{$J/\psi$}}
\newcommand{\pipi}{\mbox{$\pi^+\pi^-$}}
\newcommand{\pip}{\mbox{$\pi^+$}}
\newcommand{\pim}{\mbox{$\pi^-$}}
\newcommand{\kk}{\mbox{K^+K^-$}}
\newcommand{\bsl}{\mbox{$b$}}
\newcommand{\alp}{\mbox{$\alpha^\prime$}}
\newcommand{\alpom}{\mbox{$\alpha_{\PO}$}}
\newcommand{\alregg}{\mbox{$\alpha_{\regg}$}}
\newcommand{\alpomp}{\mbox{$\alpha_{\PO}^\prime$}}
\newcommand{\rzzzz}{\mbox{$r_{00}^{04}$}}
\newcommand{\rzqzz}{\mbox{$r_{00}^{04}$}}
\newcommand{\rzquz}{\mbox{$r_{10}^{04}$}}
\newcommand{\rzqumu}{\mbox{$r_{1-1}^{04}$}}
\newcommand{\ruuu}{\mbox{$r_{11}^{1}$}}
\newcommand{\ruzz}{\mbox{$r_{00}^{1}$}}
\newcommand{\ruuz}{\mbox{$r_{10}^{1}$}}
\newcommand{\ruumu}{\mbox{$r_{1-1}^{1}$}}
\newcommand{\rduz}{\mbox{$r_{10}^{2}$}}
\newcommand{\rdumu}{\mbox{$r_{1-1}^{2}$}}
\newcommand{\rcuu}{\mbox{$r_{11}^{5}$}}
\newcommand{\rczz}{\mbox{$r_{00}^{5}$}}
\newcommand{\rcuz}{\mbox{$r_{10}^{5}$}}
\newcommand{\rcumu}{\mbox{$r_{1-1}^{5}$}}
\newcommand{\rsuz}{\mbox{$r_{10}^{6}$}}
\newcommand{\rsumu}{\mbox{$r_{1-1}^{6}$}}
\newcommand{\rzqik}{\mbox{$r_{ik}^{04}$}}
\newcommand{\rhzik}{\mbox{$\rh_{ik}^{0}$}}
\newcommand{\rhqik}{\mbox{$\rh_{ik}^{4}$}}
\newcommand{\rhaik}{\mbox{$\rh_{ik}^{\alpha}$}}
\newcommand{\rhzzz}{\mbox{$\rh_{00}^{0}$}}
\newcommand{\rhqzz}{\mbox{$\rh_{00}^{4}$}}
\newcommand{\raik}{\mbox{$r_{ik}^{\alpha}$}}
\newcommand{\razz}{\mbox{$r_{00}^{\alpha}$}}
\newcommand{\rauz}{\mbox{$r_{10}^{\alpha}$}}
\newcommand{\raumu}{\mbox{$r_{1-1}^{\alpha}$}}

\newcommand{\R}{\mbox{$R$}}
\newcommand{\rzero}{\mbox{$r_{00}^{04}$}}
\newcommand{\rone}{\mbox{$r_{1-1}^{1}$}}
\newcommand{\costh}{\mbox{$\cos\theta$}}
\newcommand{\cosp}{\mbox{$\cos\psi$}}
\newcommand{\costop}{\mbox{$\cos(2\psi)$}}
\newcommand{\cosd}{\mbox{$\cos\delta$}}
\newcommand{\cossqp}{\mbox{$\cos^2\psi$}}
\newcommand{\cossqt}{\mbox{$\cos^2\theta^*$}}
\newcommand{\sint}{\mbox{$\sin\theta^*$}}
\newcommand{\sintot}{\mbox{$\sin(2\theta^*)$}}
\newcommand{\sinsqt}{\mbox{$\sin^2\theta^*$}}
\newcommand{\costhst}{\mbox{$\cos\theta^*$}}
\newcommand{\vep}{\mbox{$V p$}}
\newcommand{\mpipi}{\mbox{$m_{\pi^+\pi^-}$}}
\newcommand{\mkk}{\mbox{$m_{KK}$}}
\newcommand{\mkaka}{\mbox{$m_{K^+K^-}$}}
\newcommand{\mpp}{\mbox{$m_{\pi\pi}$}}       
\newcommand{\mppsq}{\mbox{$m_{\pi\pi}^2$}}   
\newcommand{\mpi}{\mbox{$m_{\pi}$}}          
\newcommand{\mrho}{\mbox{$m_{\rho}$}}        
\newcommand{\mrhosq}{\mbox{$m_{\rho}^2$}}    
\newcommand{\Gmpp}{\mbox{$\Gamma (\mpp)$}}   
\newcommand{\Gmppsq}{\mbox{$\Gamma^2(\mpp)$}}
\newcommand{\Grho}{\mbox{$\Gamma_{\rho}$}}   
\newcommand{\grho}{\mbox{$\Gamma_{\rho}$}}   
\newcommand{\Grhosq}{\mbox{$\Gamma_{\rho}^2$}}   
%
%
\newcommand{\cm}{\mbox{\rm cm}}
\newcommand{\GeV}{\mbox{\rm GeV}}
\newcommand{\gev}{\mbox{\rm GeV}}
\newcommand{\GeVx}{\rm GeV}
\newcommand{\gevx}{\rm GeV}
\newcommand{\GeVc}{\rm GeV/c}
\newcommand{\gevc}{\rm GeV/c}
\newcommand{\MeVc}{\rm MeV/c}
\newcommand{\mevc}{\rm MeV/c}
\newcommand{\MeV}{\mbox{\rm MeV}}
\newcommand{\mev}{\mbox{\rm MeV}}
\newcommand{\MeVx}{\mbox{\rm MeV}}
\newcommand{\mevx}{\mbox{\rm MeV}}
\newcommand{\GeVsq}{\mbox{${\rm GeV}^2$}}
\newcommand{\gevsq}{\mbox{${\rm GeV}^2$}}
\newcommand{\gevsqc}{\mbox{${\rm GeV^2/c^4}$}}
\newcommand{\gevcsq}{\mbox{${\rm GeV/c^2}$}}
\newcommand{\mevcsq}{\mbox{${\rm MeV/c^2}$}}
\newcommand{\GeVsqm}{\mbox{${\rm GeV}^{-2}$}}
\newcommand{\gevsqm}{\mbox{${\rm GeV}^{-2}$}}
\newcommand{\nb}{\mbox{${\rm nb}$}}
\newcommand{\nbinv}{\mbox{${\rm nb^{-1}}$}}
\newcommand{\pbinv}{\mbox{${\rm pb^{-1}}$}}
\newcommand{\mm}{\mbox{$\cdot 10^{-2}$}}
\newcommand{\mmm}{\mbox{$\cdot 10^{-3}$}}
\newcommand{\mmmm}{\mbox{$\cdot 10^{-4}$}}
\newcommand{\degr}{\mbox{$^{\circ}$}}
%
%
\newcommand{\F}{$ F_{2}(x,Q^2)\,$}  
\newcommand{\Fc}{$ F_{2}\,$}    
\newcommand{\XP}{x_{{I\!\!P}/{p}}}       
\newcommand{\TOSS}{x_{{i}/{\PO}}}        
\newcommand{\un}[1]{\mbox{\rm #1}} 
\newcommand{\LO}{Leading Order}
\newcommand{\NLO}{Next to Leading Order}
\newcommand{\ft}{$ F_{2}\,$}
%
%
\newcommand{\mc}{\multicolumn}
\newcommand{\bce}{\begin{center}}
\newcommand{\ece}{\end{center}}
\newcommand{\beq}{\begin{equation}}
\newcommand{\eeq}{\end{equation}}
\newcommand{\bea}{\begin{eqnarray}}
\newcommand{\eea}{\end{eqnarray}}
%
%
\def\lsim{\mathrel{\rlap{\lower4pt\hbox{\hskip1pt$\sim$}}
    \raise1pt\hbox{$<$}}}         
\def\gsim{\mathrel{\rlap{\lower4pt\hbox{\hskip1pt$\sim$}}
    \raise1pt\hbox{$>$}}}         
%
%
\newcommand{\pom}{{I\!\!P}}
\newcommand{\regg}{{I\!\!R}}
\newcommand{\PO}{I\!\!P}
\newcommand{\slowpi}{\pi_{\mathit{slow}}}
\newcommand{\fiidiii}{F_2^{D(3)}}
\newcommand{\fiidiiiarg}{F_2^{D(3)}\,(\beta,\,Q^2,\,x)}
\newcommand{\fiidiiifull}{F_2^{D(3)}\,(x_{I\!\!P},\,\beta,\,Q^2)}
\newcommand{\n}{1.19\pm 0.06 (stat.) \pm0.07 (syst.)}
\newcommand{\nz}{1.30\pm 0.08 (stat.)^{+0.08}_{-0.14} (syst.)}
\newcommand{\fiidiiiifull}{$F_2^{D(4)}\,(\beta,\,Q^2,\,x,\,t)$}
\newcommand{\fiipom}{\tilde F_2^D}
\newcommand{\fiipomfull}{\tilde F_2^D\,(\beta,\,Q^2)}
\newcommand{\ALPHA}{1.10\pm0.03 (stat.) \pm0.04 (syst.)}
\newcommand{\ALPHAZ}{1.15\pm0.04 (stat.)^{+0.04}_{-0.07} (syst.)}
\newcommand{\fiipomarg}{\fiipom\,(\beta,\,Q^2)}
\newcommand{\pomflux}{f_{\pom / p}}
\newcommand{\nxpom}{1.19\pm 0.06 (stat.) \pm0.07 (syst.)}
\newcommand {\gapprox}
   {\raisebox{-0.7ex}{$\stackrel {\textstyle>}{\sim}$}}
\newcommand {\lapprox}
   {\raisebox{-0.7ex}{$\stackrel {\textstyle<}{\sim}$}}
\newcommand{\pomfluxarg}{f_{\pom / p}\,(x_\pom)}
\newcommand{\dsf}{\mbox{$F_2^{D(3)}$}}
\newcommand{\dsfva}{\mbox{$F_2^{D(3)}(\beta,Q^2,x_{I\!\!P})$}}
\newcommand{\dsfvb}{\mbox{$F_2^{D(3)}(\beta,Q^2,x)$}}
\newcommand{\dsfpom}{$F_2^{I\!\!P}$}
\newcommand{\gap}{\stackrel{>}{\sim}}
\newcommand{\lap}{\stackrel{<}{\sim}}
\newcommand{\fem}{$F_2^{em}$}
\newcommand{\tsnmp}{$\tilde{\sigma}_{NC}(e^{\mp})$}
\newcommand{\tsnm}{$\tilde{\sigma}_{NC}(e^-)$}
\newcommand{\tsnp}{$\tilde{\sigma}_{NC}(e^+)$}
\newcommand{\st}{$\star$}
\newcommand{\sst}{$\star \star$}
\newcommand{\ssst}{$\star \star \star$}
\newcommand{\sssst}{$\star \star \star \star$}
\newcommand{\tw}{\theta_W}
\newcommand{\sw}{\sin{\theta_W}}
\newcommand{\cw}{\cos{\theta_W}}
\newcommand{\sww}{\sin^2{\theta_W}}
\newcommand{\cww}{\cos^2{\theta_W}}
\newcommand{\trm}{m_{\perp}}
\newcommand{\trp}{p_{\perp}}
\newcommand{\trmm}{m_{\perp}^2}
\newcommand{\trpp}{p_{\perp}^2}
%
%
\newcommand{\sqrts}{$\sqrt{s}$}
\newcommand{\Oa}{$O(\alpha_s)$}
\newcommand{\Oaa}{$O(\alpha_s^2)$}
\newcommand{\PT}{p_{\perp}}
\newcommand{\sh}{\hat{s}}
\newcommand{\uh}{\hat{u}}
\newcommand{\ttbs}{\char'134}
\newcommand{\xpomlo}{3\times10^{-4}}
\newcommand{\xpomup}{0.05}
\newcommand{\llq}{$\alpha_s \ln{(\qsq / \Lambda_{QCD}^2)}$}
\newcommand{\llqx}{$\alpha_s \ln{(\qsq / \Lambda_{QCD}^2)} \ln{(1/x)}$}
\newcommand{\llx}{$\alpha_s \ln{(1/x)}$}
%
%
%
%
\def\ar#1#2#3   {{\em Ann. Rev. Nucl. Part. Sci.} {\bf#1} (#2) #3}
\def\epj#1#2#3  {{\em Eur. Phys. J.} {\bf#1} (#2) #3}
\def\err#1#2#3  {{\it Erratum} {\bf#1} (#2) #3}
\def\ib#1#2#3   {{\it ibid.} {\bf#1} (#2) #3}
\def\ijmp#1#2#3 {{\em Int. J. Mod. Phys.} {\bf#1} (#2) #3}
\def\jetp#1#2#3 {{\em JETP Lett.} {\bf#1} (#2) #3}
\def\mpl#1#2#3  {{\em Mod. Phys. Lett.} {\bf#1} (#2) #3}
\def\nim#1#2#3  {{\em Nucl. Instr. Meth.} {\bf#1} (#2) #3}
\def\nc#1#2#3   {{\em Nuovo Cim.} {\bf#1} (#2) #3}
\def\np#1#2#3   {{\em Nucl. Phys.} {\bf#1} (#2) #3}
\def\pl#1#2#3   {{\em Phys. Lett.} {\bf#1} (#2) #3}
\def\prep#1#2#3 {{\em Phys. Rep.} {\bf#1} (#2) #3}
\def\prev#1#2#3 {{\em Phys. Rev.} {\bf#1} (#2) #3}
\def\prl#1#2#3  {{\em Phys. Rev. Lett.} {\bf#1} (#2) #3}
\def\ptp#1#2#3  {{\em Prog. Th. Phys.} {\bf#1} (#2) #3}
\def\rmp#1#2#3  {{\em Rev. Mod. Phys.} {\bf#1} (#2) #3}
\def\rpp#1#2#3  {{\em Rep. Prog. Phys.} {\bf#1} (#2) #3}
\def\sjnp#1#2#3 {{\em Sov. J. Nucl. Phys.} {\bf#1} (#2) #3}
\def\spj#1#2#3  {{\em Sov. Phys. JEPT} {\bf#1} (#2) #3}
\def\zp#1#2#3   {{\em Zeit. Phys.} {\bf#1} (#2) #3}
%
%
\newcommand{\clearemptydoublepage}{\newpage{\pagestyle{empty}\cleardoublepage}}
\newcommand{\scaption}[1]{\caption{\protect{\footnotesize  #1}}}
\newcommand{\proc}[2]{\mbox{$ #1 \rightarrow #2 $}}
\newcommand{\average}[1]{\mbox{$ \langle #1 \rangle $}}
\newcommand{\av}[1]{\mbox{$ \langle #1 \rangle $}}


\title{Diffraction at HERA and Prospects with H1
    \thanks{Talk given at the Workshop DIFFRACTION 2000, Cetraro, Italy,
    Septembre 2000.}
    }

\author{P. Marage\address{Universit\'e Libre de Bruxelles - 
    CP 230, Boulevard du Triomphe,
    B-1050 Bruxelles, Belgium \\[3pt]
    e-mail: {\tt pmarage@ulb.ac.be}}}


\begin{abstract}
After the tremendous progress achieved  at HERA in the field of 
diffraction, a new level of statistical and systematic precision 
is needed.
H1 will install a very forward proton spectrometer
in the cold section of HERA, with $\simeq 100$ \% acceptance down 
to $\modt = 0$ for $\xpom =  0.01$.
Large statistics of data, in particular in the presence of a
hard scale, will be collected with unambiguous proton tagging and 
a measurement of the $t$ distribution. 

\vspace{1pc}
\end{abstract}

\maketitle


\section{TREMENDOUS PROGRESS}

Over the last ten years, tremendous progress has been achieved at HERA
in the field of diffraction~\cite{halina}.\\

$\bullet$
The observation at high energy of a diffractive contribution of 8 to 10 \% of the total 
$\gamma p$ cross section in the DIS domain~\cite{observ_diffr}.\\

$\bullet$
The measurement of the energy dependence of inclusive diffraction 
in DIS~\cite{H1_94,Z_M_X}, 
harder than for hadron--hadron interactions~\cite{DL} and  
photoproduction~\cite{photoprod}, 
but softer than for inclusive $ep$ interactions. 
In the dipole model approach, this is attributed to the interplay of a soft 
component, due to photon fluctuations into large dipoles 
(partons with very asymmetric longitudinal momentum shares), 
and a hard component due to fluctuations into small dipoles 
(large $p_t$, small distance partons).\\


$\bullet$
A measurement~\cite {tslope} of the $t$ dependence of inclusive diffraction 
(${\rm d} \sigma / {\rm d}t \propto e^{b \cdot t})$, 
with $b \simeq 7$~\gevsqm, smaller than for soft diffraction 
but larger than for hard diffraction (e.g. \jpsi\ production, see below).\\

$\bullet$
The measurement of the diffractive structure function  
$F_2^{D(3)}\,(x_{I\!\!P},\,\beta,\,Q^2)$
\cite{H1_94,Z_M_X} and the extraction of gluon dominated parton distributions from 
positive scaling violations up to large $\beta$.
Of major importance is the factorisation theorem~\cite{Collins} justifying the use of 
these parton distributions into exclusive processes in DIS,
whereas factorisation is broken for hadron--hadron 
and for resolved photoproduction interactions.\\

$\bullet$
Compatibility of the data with the presence of a large higher twist longitudinal
component at high $\beta$~\cite{BEKW}.\\

$\bullet$
Possible indications of saturation effects at very low $x$
\cite{KGB,Amirim}, expected to be more important in 
diffraction than for the total cross section, because of the large gluon 
component.\\

$\bullet$
A study of diffractive dijet production 
in DIS~\cite{hadronic_FS,H1_dijets}, including a 
successful comparison of differential distributions with predictions based on the inclusive 
parton distributions,
showing that dijet production is a powerful tool to access 
directly the gluonic content of the pomeron.\\

$\bullet$
A measurement of open charm production in DIS~\cite{thompson}, another 
probe of the gluon content of the pomeron.
In spite of very limited statistics, differential distributions indicate, as for 
dijet production, the dominance of ($q \bar q g$) fluctuations in the proton or, 
equivalently, the presence of pomeron remnants not participating to the hard 
process.\\

$\bullet$
Studies of hadronic final states~\cite{hadronic_FS}, presenting marked 
differences with non-diffractive interactions but similarities to $e^+ e^-$ 
annihilation; Monte-Carlo simulations using inclusive parton distributions provide 
a good description of the data.\\

$\bullet$
For \jpsi\ photoproduction~\cite{VM,jpsi}, a short distance 
process governed by the large charm quark mass, 
the observation of a hard energy dependence reflecting the gluon distribution 
in the proton 
and a hard $t$ distribution ($b \simeq 4.5$~\gevsqm).
For $\rho$ mesons, the energy dependence appears to become harder as \qsq\ 
increases~\cite{ZEUS_rho,Ba}, and the $b$ slope decreases at high
\qsq\ towards that for \jpsi~\cite{Ba}.\\

$\bullet$
The dominance of $\sigma_L$ over $\sigma_T$ as \qsq\ increases 
in vector meson production~\cite{Ba,ZEUS_rho_ang}.\\

$\bullet$
The observation of a common production rate for vector mesons 
($\rho, \omega, \phi, J/\psi$) as a function of the variable 
$Q^2 + M_V^2$, when taking into account quark 
counting~\cite{VM_scaling}.\\

$\bullet$
The observation of deeply virtual Compton scattering 
($e + p \rightarrow e + p + \gamma$)~\cite{DVCS},
at a rate compatible with perturbative QCD calculations using 
skewed parton distributions.\\


\section{MISSING MEASUREMENTS AND EXPERIMENTAL UNCERTAINTIES}

In spite of this bright success, the accumulated statistics remain very 
limited in several important channels, especially in the presence of a 
hard scale (charm, high \qsq).
Several pieces of information are completely missing, and several short-comings 
affect the experimental measurements.

\subsection{$t$ distributions}

The measurement of the $t$ distribution, which is directly related to 
the size of the interacting dipole, is of crucial importance~\cite{Amirim}. 
Both the ZEUS and H1 detectors include forward proton spectrometers 
giving access to  
$t$ distributions, but their acceptance is only of the order of 5 \% 
and the accessible range 
$0.01 \lsim \xpom \lsim 0.03$ and $0.1 \lsim \modt \lsim 0.4$~\gevsq; 
in addition, the alignment 
of these detectors is difficult and leads to significant systematic uncertainties. 

No fully differential measurement has thus been performed of the inclusive
$F_2^{D(4)}\,(x_{I\!\!P},\,\beta,\,Q^2,\,t)$
structure function, and no measurements exist of $t$ distributions in the presence 
of a hard scale (jets, charm), except for vector mesons.

\subsection{Longitudinal cross section and higher twist}

The longitudinal cross section $F_L^D$ and its predicted higher 
twist behaviour are sensitive tests of different models of 
diffraction~\cite{Briskin}.
The standard way of performing the measurement is by using different 
beam energies, which has not happened so far at HERA. 

Access to $F_L^D$ is also possible through the distribution of the angle 
$\phi$ between the electron and the proton scattering planes, 
which is modulated by the longitudinal--transverse 
interference~\cite{NNN_fld}.

\subsection{Proton dissociation and other experimental uncertainties}

\begin{figure}[htbp]
\begin{center}
\epsfig{file=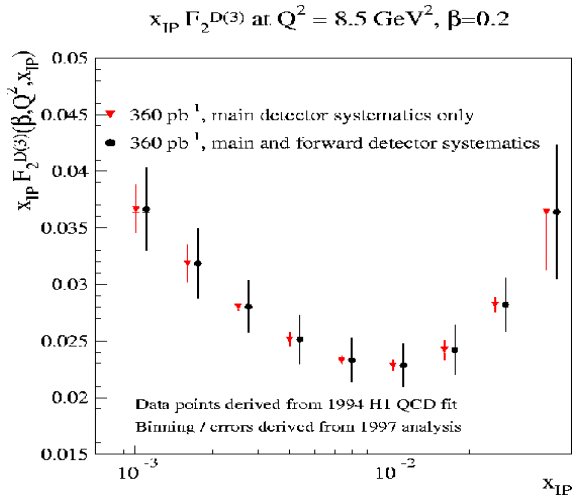,width=7.5cm,height=6.5cm}
\end{center}
\vspace{-1.0cm}
\caption {Estimated uncertainties for the measurement by H1
 of $F_2^{D(3)}$ 
 as a function of $\xpom$ in the bin ($Q^2 = 8.5$~\gevsq , $\beta = 0.2$), 
 for 350 \pbinv, the luminosity expected in 2002-2005. 
 The smaller error bars represent the errors due only to the main 
 detector measurement, the larger errors include those due to 
 the present use of the forward detectors to subtract the proton dissociation 
 background.}
\vspace{-0.5cm}
\label{fig:pdiss_error}
\end{figure}

Because of the small acceptance of the present spectrometers, most 
measurements of diffraction rely on the presence of a large rapidity 
domain devoid of hadronic energy in the region of the detector situated 
in the outgoing proton beam direction (``forward'' direction), 
without a direct tagging of the scattered proton.
This leads to large corrections, with large uncertainties, for the 
subtraction of the proton dissociation background and the extraction 
of the elastic cross section: 
$11 \pm 5$ \% for H1 and up to $31 \pm 15$ \%~\cite{Z_M_X} 
for ZEUS.
This is the dominant systematic error for many channels 
(see fig.~\ref{fig:pdiss_error}).

The proton dissociation background is also usually assumed to lead 
to an overall normalisation effect, which is certainly not justified 
(see e.g. the strong \qsq\ dependence for the ratio of proton 
dissociation to elastic $\rho$ production~\cite{drutskoi}).

Discrepancies appear in some measurements between H1 and ZEUS,
in particular for some ($Q^2, \beta $) bins of the $F_2^{D(3)}$  
structure function.
These discrepancies may disappear with new, more precise measurements. 
They may also be related to different ways of extracting the 
diffractive signal, since none of the two experiments uses the direct 
measurement of the scattered proton: 
H1 selects events with a rapidity gap~\cite{H1_94}
and ZEUS selects the non-exponentially suppressed part of the 
$M_X$ distribution~\cite{Z_M_X}.

\section{A VERY FORWARD PROTON SPECTROMETER FOR H1}

\begin{figure}[htbp]
\begin{center}
 \epsfig{file=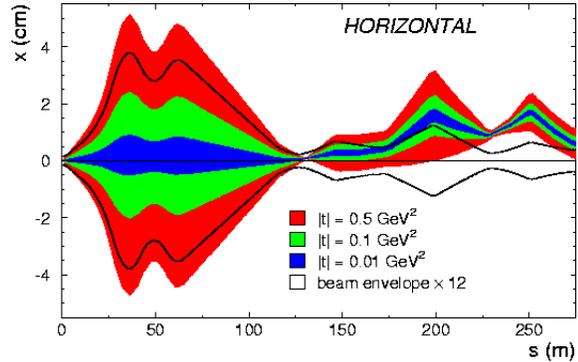,width=7.5cm}
\end{center}
\vspace{-1.0cm}
\caption {Horizontal projection of the distance of the scattered proton
  to the beam, shown for $\xpom=0.01$ and for
  three different values of $t$ (shaded areas), and 12 sigma beam envelope,
  as a function of the distance to the interaction point.
  }
\label{fig:optics}
\vspace{-0.5cm}
\end{figure}

\subsection {A very large acceptance detector}

Higher precision measurements, both from a statistical and a systematic
point of view, 
and the access to new physical quantities are necessary 
to discriminate 
between several theoretical models, which are broadly compatible with 
the present measurements. 
This implies collecting large statistics of high quality data, 
in various inclusive, semi-inclusive and exclusive channels.

The H1 collaboration has thus proposed~\cite{proposal} to install
in 2002 a very forward proton spectrometer (VFPS), consisting 
of two roman pots equipped with scintillating fiber detectors, 
in the cold section of the HERA beam line, 220~m downstream of 
the interaction point. This is where the 
spectrometer effect of the horizontal HERA bend is strongest 
(see fig.~\ref{fig:optics}).

\begin{figure}[htbp]
\begin{center}
\epsfig{file=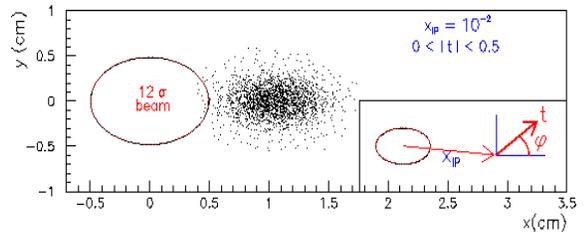,width=7.5cm,height=3.0cm}
\end{center}
\vspace{-1.0cm}
\caption {Spread of the impact point of diffractively scattered 
  protons in the 
  $(x,\,y)$ plane for $\xpom = 0.01$ and $0 \leq \modt \leq 0.5$~\gevsq, 
  with the 12 $\sigma$ beam envelope. 
  Insert: Illustration of the relation between the impact point of the 
  scattered proton in the transverse $(x,\,y)$ plane and the
  variables $\xpom$, $\modt$ and $\phi$.}
\label{fig:spread}
\vspace{-0.5cm}
\end{figure}

\begin{figure}[htbp]
\begin{center}
 \epsfig{file=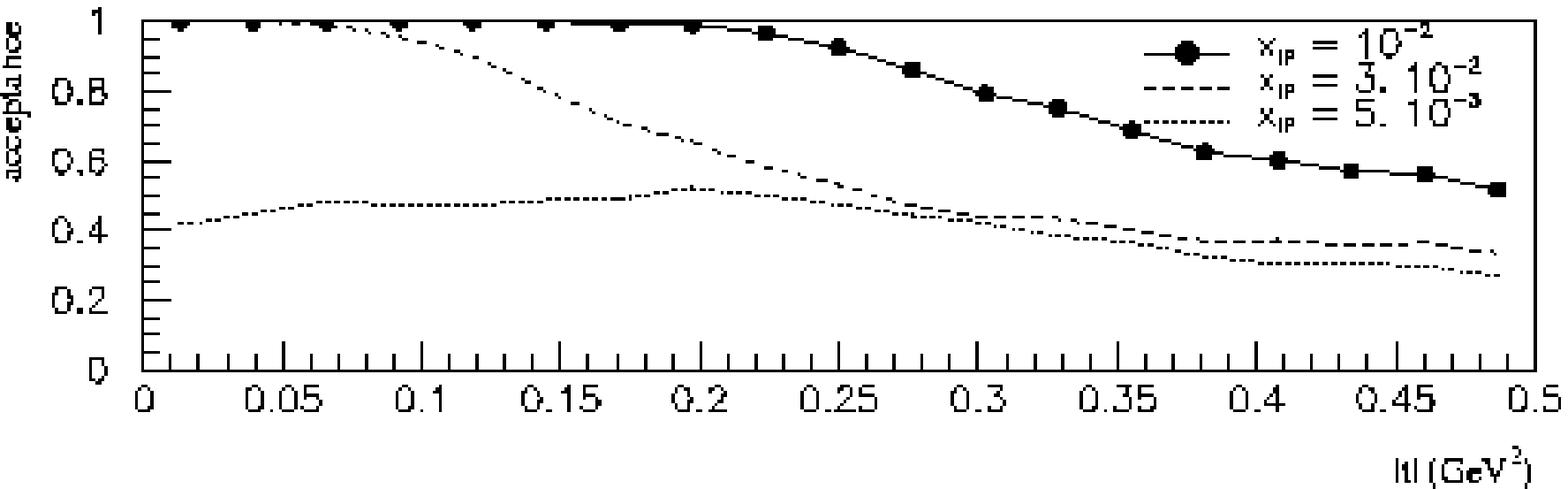,width=7.5cm,height=2.5cm}
\end{center}
\vspace{-0.5cm}
\begin{center}
 \epsfig{file=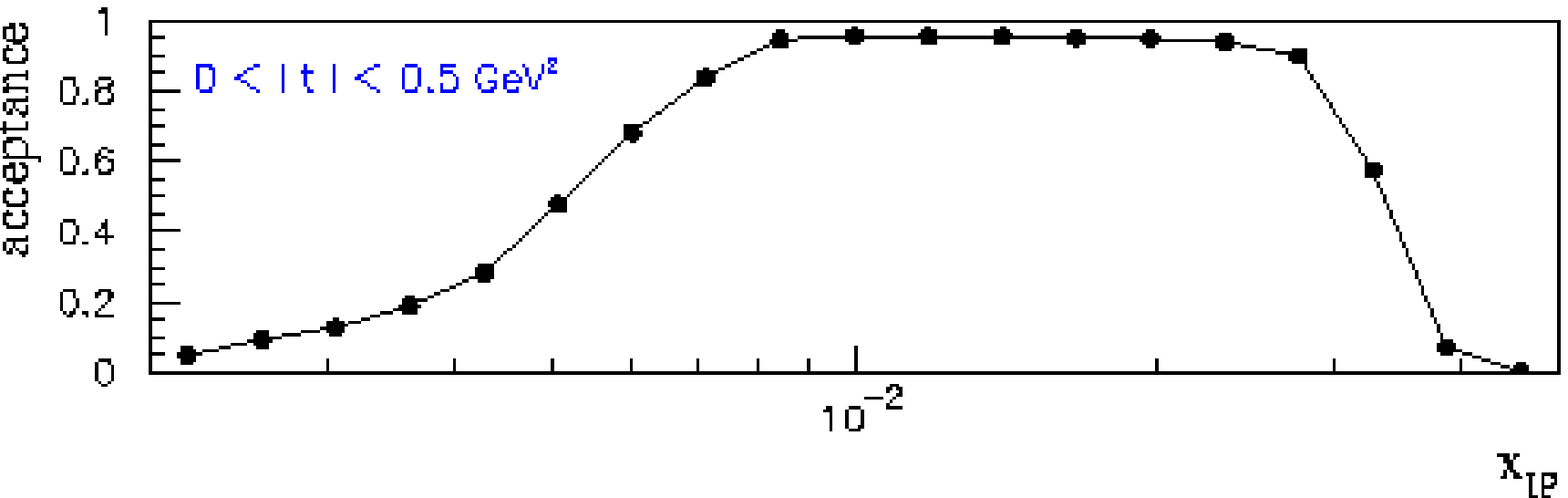,width=7.5cm,height=2.5cm}
\end{center}
\vspace{-1.0cm}
\caption {VFPS acceptances, as a function of $t$ (top) and 
of $\xpom$ (bottom).}
\label{fig:acc}
\vspace{-0.5cm}
\end{figure}

\begin{figure}[htbp]
\begin{center}
 \epsfig{file=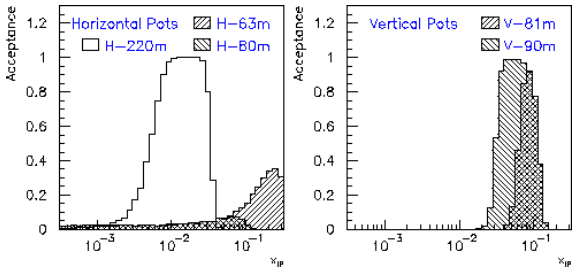,width=7.5cm}
\end{center}
\vspace{-0.8cm}
\caption {Acceptance as a function of $\xpom$ for the VFPS and the 
present H1 spectrometer (horizontal and vertical stations, 
respectively).}
\label{fig:compa}
\vspace{-0.5cm}
\end{figure}

The acceptance will be $\simeq 100 \% $ for $\xpom = 0.01$
(see figs.~\ref{fig:spread} and~\ref{fig:acc}), down to the lowest 
$\modt$ values.
This is a large improvement compared to the present proton
spectrometers (see fig.~\ref{fig:compa}).

Full advantage will thus be taken of the 5-fold luminosity increase 
after the HERA upgrade, and uncertainties in the cross
section measurements due to extrapolations in $t$ are avoided.
Also, the error due to the uncertainty on the precise positioning of 
the pots is avoided, in contrast with the present case where the
acceptance varies strongly with $t$.
The detector will be aligned using the position of the $t$ peak and 
$\rho$ photoproduction.

\subsection{Physics issues}

With the VFPS, the direct tagging of the scattered proton will 
eliminate the largest source of uncertainty in most channels, i.e.
the proton dissociation background.

The emphasis will be put on precise measurements of diffractive processes
in the presence of a hard scale: high \qsq\ values, large transverse
momenta (large $E_T$ jets), large quark masses (charm production), and on
the measurement of the $t$ slope.

For the inclusive and the semi-inclusive processes (jets, charm production), 
the measurement of the $t$ distribution with 3 to 4 bins for 
$0 \leq \modt \leq 0.6$~\gevsq\ will permit discriminating between
large ($b \gsim\ 7$~\gevsqm) and small  
($b \simeq 4.5$~\gevsqm) dipoles.\\

$\bullet$
The fully differential structure function $F_2^{D(4)}$ will be measured
in the purely diffractive domain $\xpom \lsim 0.01$, where the
contribution from meson exchange is negligible, in 3 or 4 bins in~$t$.
The $\xpom$ variable will be measured both with the VFPS and using the main
detector, leading to improved precision.\\

$\bullet$
For 350 $\pbinv$ data taken in 2002-2005, 10~000 dijet events with 
$p_T > 5 $~\gev,
$Q^2 > 7.5$~\gevsq\ and $0.1 < y < 0.7$ are expected to be tagged 
by the VFPS.
It will be possible to estimate higher order contributions, like 
resolved virtual photon.
In photoproduction, 60~000 events are expected, and comparison between 
rates for direct ($x_\gamma > 0.8$) and resolved ($x_\gamma < 0.8$) 
photons will permit investigating the mechanisms which destroy the
rapidity gap (``gap survival probability''~\cite{survival}).\\

$\bullet$
Some 400 charm events are expected in the truly 
diffractive domain $5 \cdot 10^{-3} < \xpom < 3 \cdot 10^{-2}$, 
to be compared with a few tens of events presently accumulated 
on basis of the presence of a rapidity gap.
Given the statistical precision (5~\%), it is essential for a cross
section measurement to tag the scattered proton and to eliminate 
the systematic error due to proton dissociation.\\

$\bullet$
Large samples of high \qsq\ vector meson and DVCS events will be
accumulated (7000 DVCS events with $Q^2 > 8$~\gevsq), with no proton
dissociation background.\\

$\bullet$
Information will be obtained on the longitudinal structure function 
using the modulation of the diffractive cross section in 
$\phi$, the angle between the electron and proton scattering 
planes. 
According to models, the effect can be of the order of 
$15 - 25$~\% for $\beta \gsim\ 0.8$, depending on \qsq\ and $\beta$.
The angle $\phi$ (see fig.~\ref{fig:spread}) will be measured in 
4 or 5 bins, for $\modt > 0.1$~\gevsq.\\

$\bullet$
Proton dissociation samples of events will be selected as the difference
between the events selected by the observation of a rapidity gap in the
main detector, and those which are in addition tagged by the VFPS.
This will open a completely new field in diffraction, related to
factorisation breaking effects expected in QCD.\\


\section{CONCLUSIONS}
With the increase of luminosity after 2001 and the installation 
by H1 of the VFPS in 2002, very significant improvements will be achieved
in HERA to the precision of the diffraction measurements.
The dipole approach will be quantitatively tested, discrimination between
several models will be possible, and a new field of research will be
opened with the study of factorisation breaking.


\section*{ACKNOWLEDGEMENTS}
It is a pleasure to thank the organisors of DIFFRACTION 2000, and in
particular R. Fiore, for a very pleasant and fruitful workshop.



\end{document}